\documentclass[12pt,preprint]{aastex}
\usepackage{cite}
\usepackage{graphicx}
\usepackage{geometry}
\usepackage[nooneline,flushleft]{caption2}
\usepackage{natbib}
\usepackage{hyperref}  % Adding hyper refs (green squre links to refs)

\shorttitle{Low Surface Brightness Galaxies}
\shortauthors{Cao et al.}
\begin{document}
\title{Molecular Gas and Star-Formation In Low Surface Brightness Galaxies}
\author{Tian-Wen Cao$^{1,2,3}$, Hong Wu$^*$ $^{1,2}$, Wei Du$^{1}$, Feng-Jie Lei$^{1,2}$, Ming Zhu$^{1}$, Jan Wouterloot$^{4}$, Harriet Parsons$^{4}$, Yi-Nan Zhu$^{1}$, Chao-Jian Wu$^{1}$, Fan Yang$^{1}$, Chen Cao$^{5}$, Zhi-Min Zhou$^{1}$, Min He$^{1,2}$, Jun-Jie Jin$^{1,2}$, James E. Wicker$^{1}$}

\altaffiltext{1}{Key Laboratory of Optical Astronomy, National Astronomical Observatories, Chinese Academy of Sciences, Beijing 100012, P.R. China; twcao@bao.ac.cn}
\altaffiltext{2}{School of Astronomy and Space Science, University of Chinese Academy of Sciences, Beijing, P.R. China; hwu@bao.ac.cn}
\altaffiltext{3}{Chinese Academy of Sciences South America Center for Astronomy, China-Chile Joint Center for Astronomy,Camino El Observatorio 1515, Las Condes, Santiago, Chile}
\altaffiltext{4}{East Asian Observatory, 660 N. A‘oh$\overline{o}$k$\overline{u}$ Place, University Park, Hilo, Hawaii 96720, USA}
\altaffiltext{5}{School of Space Science and Physics, Shandong University at Weihai, Weihai, Shandong 264209, China}

\begin{abstract}
We have obtained CO(J=2-1) spectra of nine face-on low surface
brightness galaxies(LSBGs) using the JCMT 15-meter telescope and
observed $H\alpha$ images using the 2.16-meter telescope of NAOC. As no CO has been detected, only upper limits on the H$_2$ masses are given. The upper limits of total molecular hydrogen masses are about (1.2-82.4) $\times$
10$^7\,$M$_\odot$. Their star formation rates are mainly lower than 0.4
M$_\odot\,$yr$^{-1}$ and star formation efficiencies are lower than 1.364
$\times$ 10$^{-10}\,$yr$^{-1}$. Our results show that the absence of molecular gas
content is the direct reason for the low star formation rate. The
low star formation efficiency probably resulted from the low
efficiency of HI gas transforming to H$_2$ gas.
\end{abstract}

\section{Introduction \label{intro}}

Low Surface Brightness galaxies (LSBGs,\citet{1997ARA&A..35..267I}) are important for investigating the evolution of our universe. The origin and evolution of these LSBGs
are still mysterious. They have significantly different chemical
enrichment histories from normal galaxies
\citep{2001ApJ...557..495P,2011MNRAS.417.1335P}. LSBGs are usually
optically faint and blue\citep{1995MNRAS.274..235D,1997ARA&A..35..267I}. The stellar disks of most LSBGs are diffuse. They usually have low
metallicities (Z$<$1/3$\,$Z$_\odot$ \citet{1994ApJ...426..135M}), low column densities (N$_{HI}\sim$10$^{20}\,$cm$^{-2}$ \citet{1996MNRAS.283...18D}) and low dust masses\citep{2001ApJ...548..150M}. The star formation rates (SFRs) of
LSBGs are lower than those of normal
galaxies\citep{1999A&A...342..655G,2008ApJ...681..244B,2009ApJ...696.1834W}.

According to pervious works, the HI content of most LSBGs is rich,
compared with normal star-forming galaxies
\citep{1994ApJ...426..135M}. The HI gas disk extends well beyond the stellar disk
\citep{1997ApJ...481..689M,1999A&A...342..655G,2004A&A...428..823O,2001A&A...365....1M}, and is about double the size of
the optical disk\citep{1996MNRAS.283...18D,1997AJ....114.1858P,2007MNRAS.379...11D}.

Although the original material involved in star formation is HI, the
star formation is indirectly related with HI. Generally, the
star formation arises out of molecular clouds. The low star
formation rates in LSBGs may be related to the absence of molecular
gas content. The optical peculiarities of LSBGs have been
discussed by some works \citep{1993AJ....106..548V, 1994AJ....107..530M, 1994ApJ...426..135M, 1995MNRAS.274..235D, 1997ARA&A..35..267I, 1998MNRAS.299..123J}, however, the cold molecular gas in these galaxies is far from well understood. Molecular gas is vital for studying the star formation process.
There are some previous works which try to detect the CO content in LSBGs\citep{1990AJ....100.1523S,1993PhDT.........2K,1998A&A...336...49D,2000A&A...358..494B}.
Most works just give upper limits on CO content, and only a few LSBGs
have detected molecular gas
\citep{2001ApJ...549L.191M,2003ApJ...588..230O,2005AJ....129.1849M,2010A&A...523A..63D,2011AJ....142..170H}.
This may indicate a shortage of molecular gas in LSBGs.

To explore the low star formation efficiency of LSBGs, we need to
know which phase is dominant during star formation. There are two
phases in formation of a star: Firstly, the HI gas transforms into
molecular gas, then molecular gas forms a star. The CO(J=2-1)
emission line is used to trace molecular hydrogen gas in this work.
We observe the CO(J=2-1) emission line by JCMT, H$\alpha$ images by
the 2.16-meter telescope of NAOC and also combine with NUV data from
GALEX and HI data from Arecibo.

In this paper, the sample and observations of LSBGs are presented in
\textsection 2. Results and analysis are given in \textsection 3.
Discussion and summary are provided in \textsection 4 and
\textsection 5.

\section{The Sample and Observation of LSBGs \label{samXobs}}

\subsection{Sample \label{S}}
The Arecibo Legacy Fast ALFA (ALFALFA) survey
\citep{2005AJ....130.2598G}, which covers 7000$\,$deg$^2$ of high
Galactic latitude sky, provides a 21-cm HI emission line spectral
database with redshift from 1600 to 18000$\,$kms$^{-1}$ and velocity
resolution of 5$\,$kms$^{-1}$. The $\alpha$.40 catalog is the first
released catalog, which covers 40$\%$ of the area of the ALFALFA
survey and contains 15855 objects \citep{2011AJ....142..170H}. About
78$\%$ sources have optical counterparts from the Sloan Digital Sky
Survey (SDSS,\citet{2011AJ....142..170H}).

\begin{figure}[h!tb]
\captionstyle{flushleft}
\onelinecaptionstrue
\begin{center}
\includegraphics[width=8.4in]{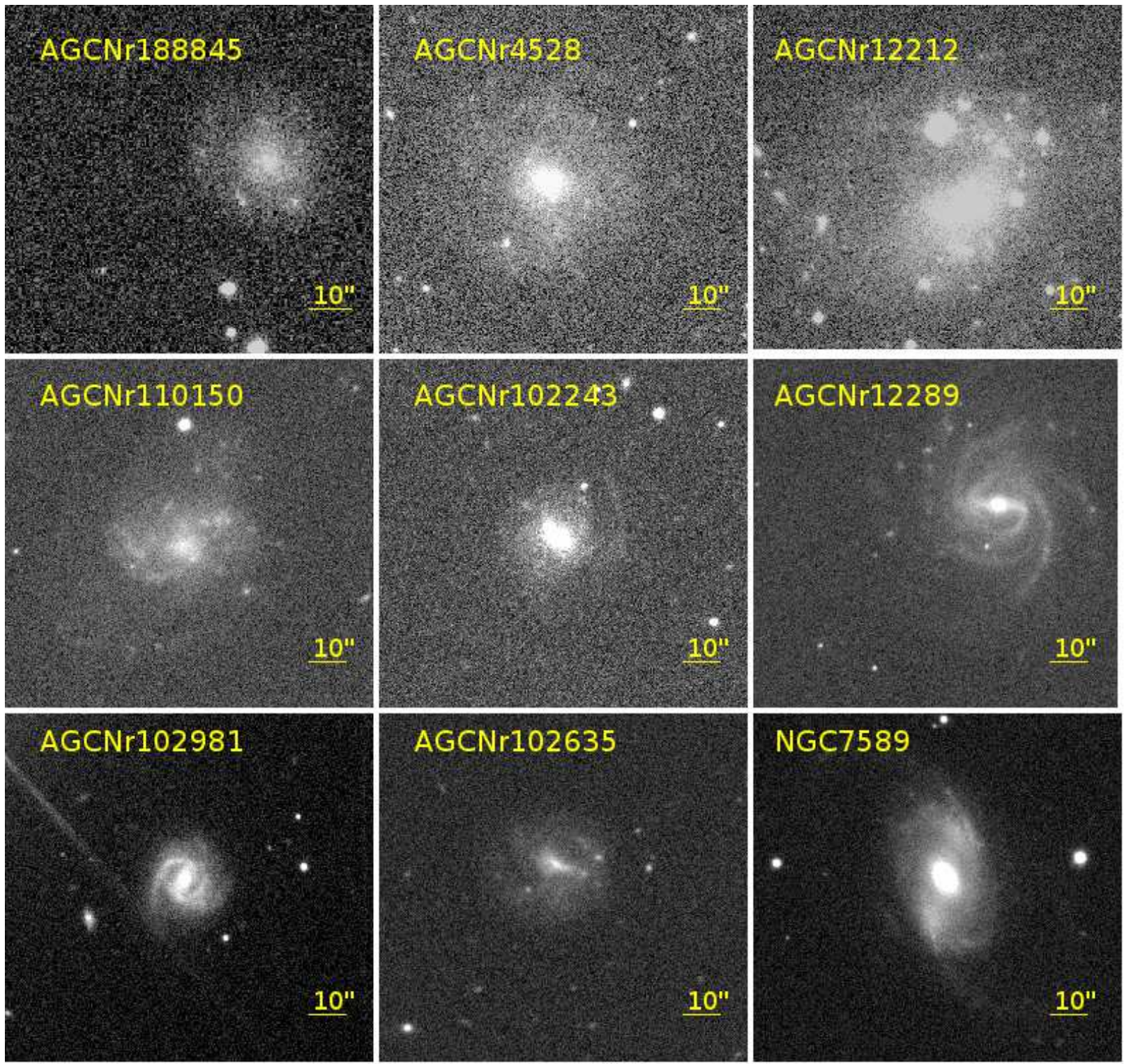}
\end{center}
\caption{SDSS r band images of our LSBGs.
       \label{rband}}
\end{figure}

Based on the $\alpha$.40 catalog, \citet{2015AJ....149..199D}
selects a sample of LSBGs which contains 1129 LSBGs with surface
brightness $\mu_{B(0)obs}$ larger than 22.5 mag/arcsec$^{2}$.
We selected nine LSBGs from this sample. 
They all have face-on disks and their HI masses are about
(0.52-19.9) $\times$ 10$^9\,$M$_\odot$, which are expected to have higher fluxes of CO(J=2-1) emission lines than others in the 1129 LSBGs sample. The redshifts in our sample are between 0.0029 and 0.0339. Generally, the size of the CO disk is about half the size of the optical disk
\citep{1989ApJ...347L..55Y} and the optical sizes of nine targets
are about 30-80 arcseconds. Figure \ref{rband} shows r-band
images of our nine LSBGs from SDSS DR12. More details about
properties of the galaxies are shown in Table \ref{tbl1}.

\subsection{CO(J=2-1) emission \label{CO}}
We obtained CO(J=2-1) spectra of nine targets using the R$\times$A3
receiver with ACSIS as the backend, mounted on the JCMT 15-meter
telescope near the peak of Mauna Kea, Hawaii. The half power beam
width (HPBW) is about 20$''$ at 230$\,$GHz. The frequency coverage of the
A3 receiver ranges from 211.5 to 276.5$\,$GHz and the CO(J=2-1) spectra
of our objects shift from 222.99 to 229.86$\,$GHz. The A3 receiver
gives 1936 channels over a bandwidth of 1$\,$GHz with a channel
separation of 0.516$\,$MHz, and the velocity resolution is 0.674$\,$km/s
(one channel).

The targets are finally observed under band 4 weather\footnotemark[6] in March and
band 5 weather in August of 2015.
In band 4 weather the atmospheric zenith opacity at 225 GHz, as measured with the Water Vapour Monitor (measured in the direction in which the telescope is observing), or the CSO tau meter (measuring in a fixed direction), is between 0.12 and 0.20. In band 5 weather the zenith opacity is between 0.20 and 0.32.
The on-source intergration time per scan was 400 seconds and one such scan took 13 minutes. The observation mode was $'$double beam switching$'$\footnotemark[7] over 60 arcseconds(to both sides with respect to the source). The total integration time of each object is about two hours and eight hours for band 4 weather and band 5 weather, respectively.
The data are calibrated by ORAC-DR\citep{2005StaUN.236.....H} data-reduction
pipeline\footnotemark[8]
using STARLINK software. In Table \ref{tbl2}, columns 2-6 show
details about the observations of CO(J=2-1) emission line.

\footnotetext[6]{http://www.eaobservatory.org/jcmt/observing/weather-bands/}

\footnotetext[7]{http://www.eaobservatory.org/jcmt/instrumentation/heterodyne/observing-modes/}

\footnotetext[8]{http://starlink.eao.hawaii.edu/devdocs/sun260.htx/sun260.html}

\subsection{H$\alpha$ Images \label{Ha}}

The H$\alpha$ images of NGC7589 and AGCNr12212 have been observed by
\citet{2000AJ....119.2757V}, \citet{2008MNRAS.390..466E} and
\citet{2016MNRAS.455.3148S}. The H$\alpha$ images of other seven
LSBGs were observed by the 2.16-meter telescope at Xinglong
Observatory, administered by the National Astronomical Observatories, Chinese Academy of Sciences (NAOC). This facility houses the BAO
Faint Object Spectrograph and Camera
(BFOSC)\citep{2016arXiv160509166F} with a 1272$\times$1152 E2V CCD. The field of view is about 9.46 arcmin $\times$ 8.77
arcmin. The pixel size is 0.475 $''$ and the gain is 1.08$\,$e/ADU.
Each target was observed with a broad band R filter and one narrow band H$\alpha$ filter.
The center wavelength of the R-band filter is 6407$\,\AA$ and the FWHM is 1580$\,\AA$.
According to the different redshifts of objects, we employed
H$\alpha$ filters with different central wavelength of 6660$\,\AA$,6710$\,\AA$ and
6760$\AA$ and FWHM of 70$\AA$. The exposure times are about 600
seconds and 1800 seconds for the R-band and H$\alpha$ narrow bands,
respectively.

The image reduction is performed by using IRAF software and the sky
background subtraction applied the more accurate method by
\citet{1999AJ....117.2757Z}, \citet{2002AJ....123.1364W} and
\citet{2015AJ....149..199D}. The stellar continuum of each Ha image is 
removed by substracting scaled R-band image. 
Finally we measured the Ha fluxes of these LSBGs, using the ellipse photometry of IRAF. 
In Table \ref{tbl2}, columns 7-10 list the details of observations of
H$\alpha$ images. AGCNr102981 is affected by light pollution from a
nearby bright star.

\section{Results and Analysis \label{RXA}}

\subsection{$H_2$ Masses \label{HM}}
\begin{figure}[h!tb]
\captionstyle{flushleft}
\onelinecaptionstrue
\begin{center}
\includegraphics[width=8.5in]{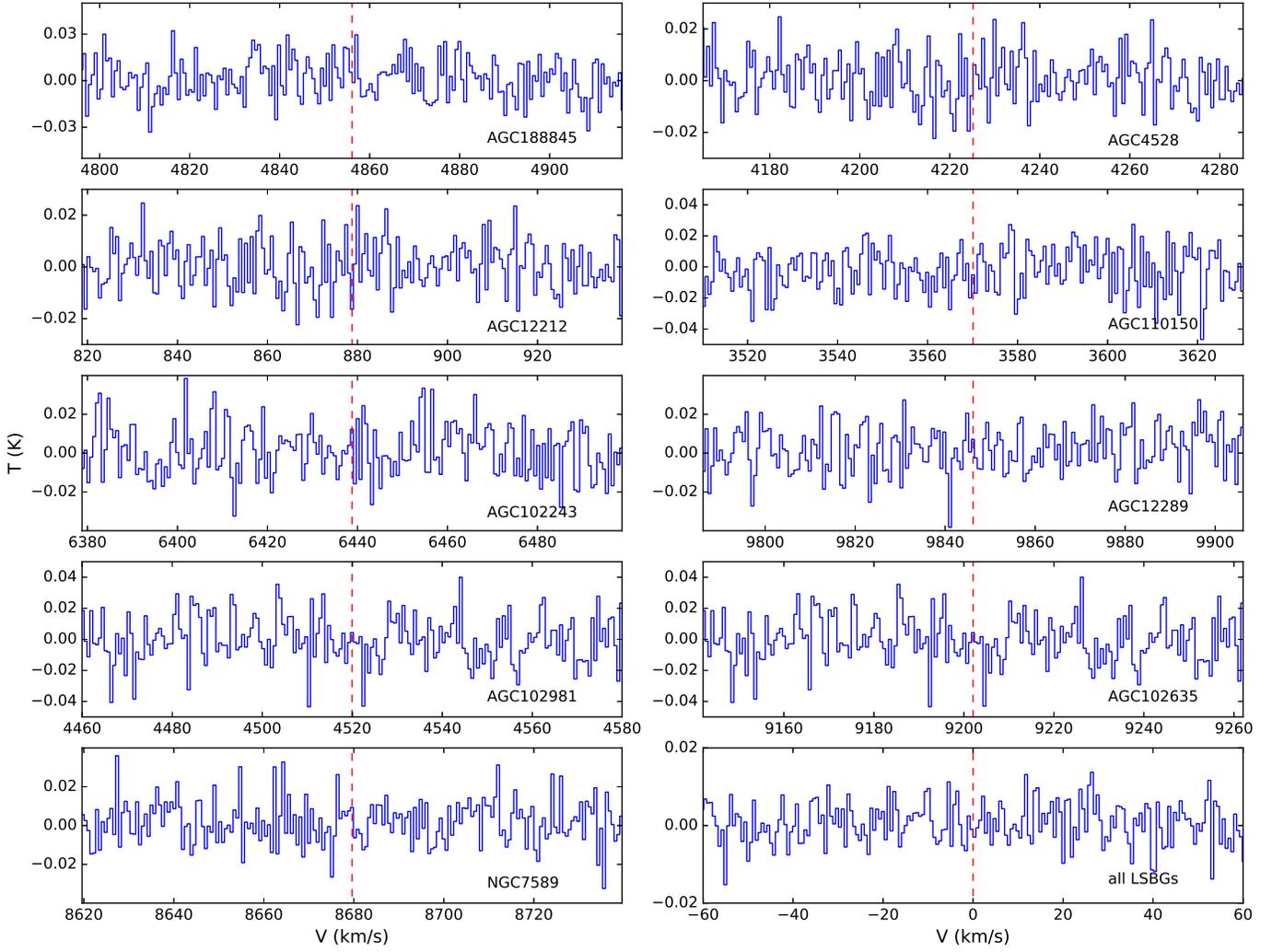}
\captionsetup{margin=80pt}\caption{Nine individual and combined CO(J=2-1) spectra of LSBGs. The red dashed line mark the position of emission line of redshfited CO(J=2-1).
       \label{spectra}}
\end{center}
\end{figure}

To enhance signal-to-noise, we binned the CO(J=2-1) spectral
channels. The final CO(J=2-1) spectral have been smoothed to 15$\,$km/s
as shown in Figure \ref{spectra}. Apparently, None of nine
targets are detected CO(J=2-1) content.

To estimate the molecular hydrogen masses, we adopted W$_{50}$ as
the linewidth. Here, W$_{50}$ is the Full Width at Half Maxinum(FWHM) of HI emission line.
The W$_{50}$ of our targets is about 65-345$\,$km/s.
The aperture efficiency $\eta _a$ is 0.61 at 225$\,$GHz, and the
conversion from T$_A$ in Kelvin to flux density in Jansky
is S(Jy) = 15.6 $\times$ T$_A$(K)/$\eta _a$, for JCMT. In this work, the T$_A$ is replaced by the 3$\sigma/\sqrt{n}$, where the $\sigma$ is the rms noise at native resolution and $n$ is the number of channels to smooth velocity resolution to 15$\,$km/s.
Hence, flux densities in Jansky are calculated according to equation:
S$_{CO(J=2-1)}$(Jy) = 15.6 $\times$ 3 $\times$
$\sigma/\sqrt{n}$/$\eta _a$. Here we adopt the R$_{21}$=CO(J=2-1)/CO(J=1-0)=0.7
\citep{2009AJ....137.4670L,2012AJ....143..138S};

The CO-to-H$_2$ conversion equation is as follows:

M$_{H_2}$ = $\alpha _{CO}$ $\times$ L$_{CO} $ $\quad $ $\quad $ $\quad $ $\quad$ $\quad$ $\quad$ $\quad$ $\quad$  $\quad$ $\quad$  $\quad$  $\quad$ $\quad$$\quad$$\quad$ $\quad$ $\quad$ $\quad$ $\quad$(1)

L$_{CO}$ = 3.25 $\times$ 10$^7$ $\times$ S$'_{CO}$ $\times$ ${\nu_{obs}}^{-2}$ $\times$ D$^2$ $\times$ (1+z)$^{-3}$ $\quad$ $\quad$  $\quad$ $\quad$ $\quad\quad\quad\quad$(2)

In Equation(1), M$_{H2}$ is molecular hydrogen mass in M$_\odot$ and
L$_{CO}$ is CO luminosity in K$\,$km$\,$s$^{-1}\,$pc$^2$. In Equation(2),
S$'_{CO}$ is integrated CO flux density in Jykms$^{-1}$. We
adopt linewidth W$_{50}$ and the resolution 15$\,$km/s, Such
S$'_{CO_{(J=1-0)}}$ = S$_{CO_{(J=1-0)}}$ $\times$ W$_{50}$/15.
${\nu _{obs}}$ is observation frequency in GHz, D is distance in Mpc
and z is redshift.

In different environments, the X$_{CO}$ factor is different.
Considering that the conversion factor of CO-to-H$_2$ increases with decreasing metallicity, X$_{CO}$ is chosen to be 3.162 $\times$ 10$^{20}\,$cm$^{-2}$/(Kkms$^{-1}$) (X$_{CO}$ = $\alpha _{CO} \times$ 6.3$\times$ 10$^{19}\,$pc$^2$cm$^{-2}$M$_\odot^{-1}$, $\alpha _{CO}$ simply is a mass-to-light ratio in M$_\odot$(K$\,$km$\,$s$^{-1}\,$pc$^2$)$^{-1}$ \citet{2013ARA&A..51..207B}) which is also the same as previous works\citep{2001ApJ...549L.191M,2005AJ....129.1849M}.
From Equation (1), we can calculate upper limits of
molecular hydrogen masses per beam, which are about
(0.68-31.6) $\times$ 10$^7\,$M$_\odot$.

To compare with previous works, we correct the beam size to the CO
disk size of our LSBGs. We estimate the upper limits of total
molecular hydrogen masses of our sample. The beam filling factors are defined as the ratios between the area of beam and the total area of CO disk which is generally half of the optical disk\citep{1989ApJ...347L..55Y}, and listed in Table 3.
The major-axis of optical disks are given by R band images and listed in the Table 1. The size of NGC7589 is from \citet{1989spce.book.....L}. The upper limits of total
molecular hydrogen masses are about (1.2-82.4) $\times$
10$^7\,$M$_\odot$ which are listed in Table \ref{tbl3}. Since the HI disk
is more extended in LSBGs, the CO disk may be larger than half of
optical disk. The total molecular hydrogen masses may be
underestimated in Table 3. In the following parts of this paper,
M$_{H_2}$ represents the total molecular hydrogen mass.

\begin{figure}[h!tb]
\captionstyle{flushleft}
\onelinecaptionstrue
\begin{center}
\includegraphics[width=8.0in]{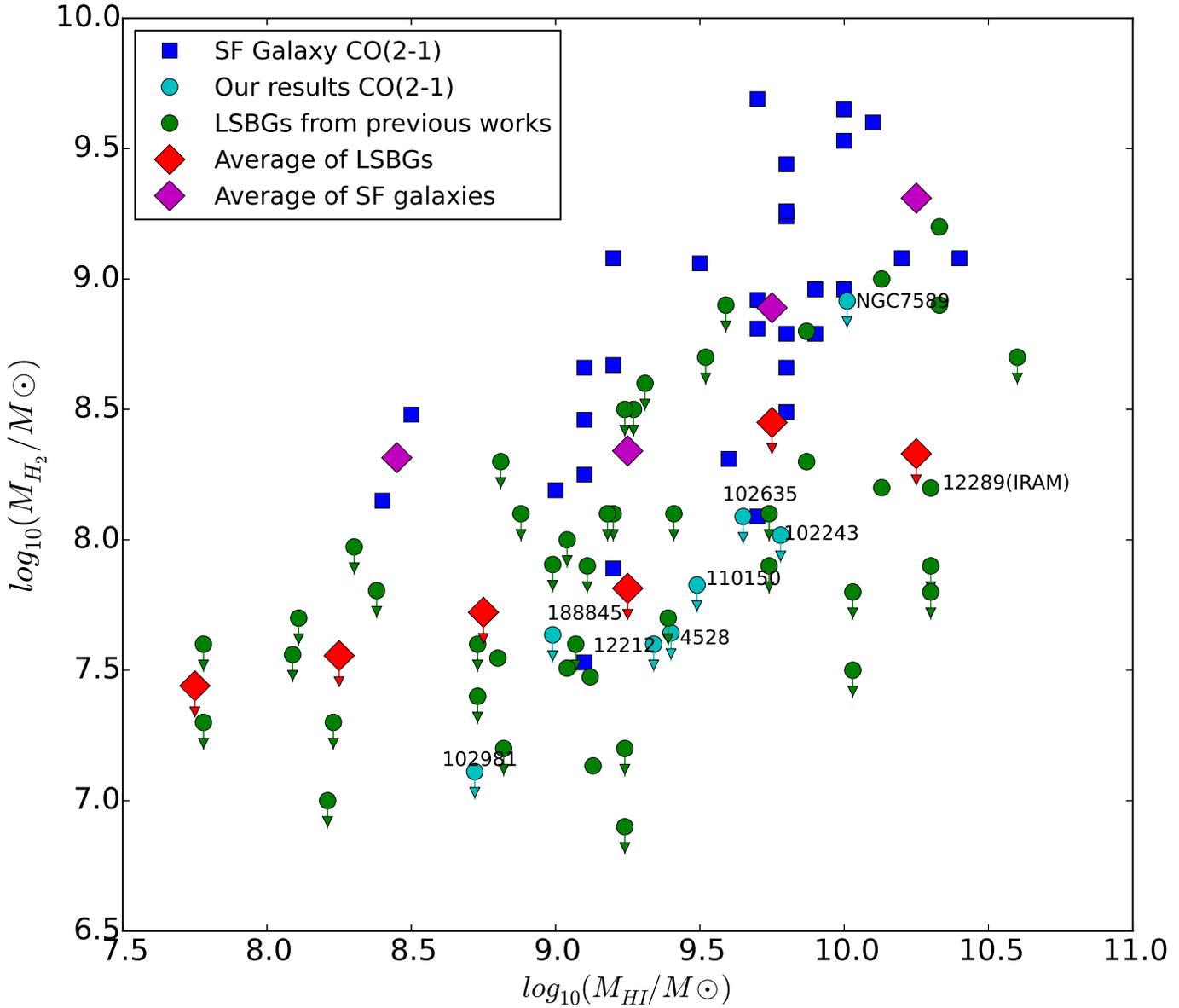}
\end{center}
\captionsetup{margin=80pt} \caption{HI mass vs. H$_2$ mass. The cyan dots are LSBGs from this survey, the green dots are LSBGs
from previous works\citep{1990AJ....100.1523S,1998A&A...335..421D,2000ApJ...545L..99O,2000A&A...358..494B,2003ApJ...588..230O}.
Blue squares are SF galaxies from \citet{2015ApJ...799...92J}.  The red and purple diamonds presents the average molecular mass of LSBGs and SF galaxies, respectively. The arrows indicate the upper limits.
       \label{result}}

\end{figure}

In Figure \ref{result}, we plot the M$_{HI}$ versus M$_{H_2}$
including our sample and LSBGs from some previous works. We have calculated the average molecular mass with bin size in 10$^{0.5}\,$M$_\odot$. The average of M$_{H_2}$ of LSBGs is smaller than those in star forming (SF) galaxies for a given HI mass.
We can see that upper limits of M$_{H_2}$ in our work are consistent with previous works. LSBGs are deficient of the molecular gas compared with these SF galaxies.

32 nearby gas-rich SF galaxies in Figure \ref{result} are from
\citet{2015ApJ...799...92J} and observed with the Sub-millimeter
Telescope (SMT).
Their HI masses are from the ALFALFA catalog. These SF galaxies are intermediate-mass galaxies ($<$10$^{10}\,$M$_\odot$). The intermediate-mass galaxies were found to be more gas rich \citep{2009ARA&A..47..159B}. These galaxies present the low-mass end of main sequence star-forming galaxies.
Both the molecular mass of our sample and SF galaxies sample are calculated
by the CO(J=2-1) emission line, so the SF galaxies are selected as a
comparison sample.

Some reasons could explain the lack of molecular gas in LSBGs.
Firstly, metallicity can affect the cooling efficiency of the
Interstellar Medium (ISM) which may impact the formation of Giant
Molecular Clouds (GMCs). Secondly, dust grains where H$_2$ forms
\citep{1979ARA&A..17...73S} can shield molecular gas from
photodissociation. The low interstellar medium (ISM) densities make
it hard for molecular clouds to form and be maintained.

The low metallicities and low ISM densities in LSBGs can make it either difficult to form H$_2$ or easy to destroy H$_2$.

\subsection{Star Formation Rates \label{SFR}}
We adopt H$\alpha$ luminosity to calculate SFR of our sample. As the
dust in LSBGs is usually less, few targets can be detected in 22
$\mu$m (WISE W4) band, so we ignore the dust extinction in
our sample.

The SFR of AGCNr12212 has been found to be 0.111$\,$M$_\odot$yr$^{-1}$ in the
work of \citet{2000AJ....119.2757V} and 0.056$\,$M$_\odot$yr$^{-1}$ in the
work of \citet{2008MNRAS.390..466E}. Six H$\alpha$ fluxes of LSBGs
are available from our observation. We transform H$\alpha$ fluxes to
luminosity using the relation: L = 4 $\times$ $\pi$ $\times$ D$^2$
$\times$ F, where D is distance in centimeter and F is H$\alpha$
flux in erg/s/cm$^{2}$. We employ the following equation to
calculate SFR\citep{1998ARA&A..36..189K}:

SFR$_{H\alpha}$(M$_\odot$yr$^{-1})$ = 7.9 $\times$ 10$^{-42}$ $\times$ L$_{H\alpha}\,$erg/s $\quad$ $\quad$  $\quad$ $\quad$ $\quad$$\quad$ $\quad$ $\quad$$\quad$ $\quad$(3)

The derived SFRs are from 0.056 to 0.829$\,$M$_\odot$yr$^{-1}$.

We also use the NUV-band luminosity to calculate SFR of four sources
(AGCNr4528, 110150, 102635, NGC 7589) which have NUV data from
GALEX. We adopt the following equation \citep{1998ARA&A..36..189K}:

SFR$_{UV}$(M$_\odot$yr$^{-1}$) = 1.4 $\times$ 10$^{-28}$ $\times$ L$_{\nu}\,$(erg/s/Hz) $\quad$ $\quad$  $\quad$ $\quad$ $\quad$$\quad$ $\quad$ $\quad$$\quad$(4)

The NUV-based SFRs are 0.146$\sim$1.003 M$_\odot$yr$^{-1}$, which are
systematically larger than those calculated from H$\alpha$
luminosity. The UV emission is also contaminated by older stars, so this would lead to deriving higher SFR from UV than that from the H$\alpha$. All these values are listed in the Table \ref{tbl3}.

In Figure \ref{SFR}, we show SFR versus HI mass for our sample and
black lines connect the same target. We also show another
sample of LSBGs in green dots from \citet{2008ApJ...681..244B} whose
SFRs are estimated by NUV luminosity. The Figure shows an increase of the SFR with HI mass. SFR in LSBGs is about 0.2$\,$M$_\odot$yr$^{-1}$ in the model calculations of \citet{1994AJ....107..530M} and 0.17-0.36$\,$M$_\odot$yr$^{-1}$ by H$\alpha$ \citep{2001AJ....122.2318B}, they are lower than that of SF galaxies. The low SFRs of our sample are consistent with previous results.
The low SFR in LSBGs agrees with the low molecular hydrogen mass.

\subsection{Star Formation Efficiency \label{SFE}}

Star formation efficiency (SFE) \citep{2008AJ....136.2782L} is the
ratio of star formation rate to total mass of gas (SFE =  SFR /
M$_{gas}$). M$_{gas}$ is the total gas mass which can be estimated by
H$_2$ mass and HI mass (M$_{H_2}$ + M$_{HI}$).

The SFR and M$_{H_2}$ of our sample have been calculated in
\textsection 3.1 and \textsection 3.2. M$_{HI}$ is from the ALFALFA
catalog listed in Table \ref{tbl2}. Since M$_{H_2}$ is far smaller
than M$_{HI}$, we adopt M$_{HI}$ to replace M$_{gas}$. SFE can be
calculated by SFR$_{H\alpha}$ and SFR$_{UV}$, respectively. SFEs of
our LSBGs are (0.165$\sim$1.364) $\times$ 10$^{-10}\,$yr$^{-1}$ by
SFR$_{H\alpha}$ and (0.471$\sim$1.174) $\times$ 10$^{-10}\,$yr$^{-1}$
by SFR$_{UV}$, respectively. The results are shown in Table \ref{tbl3}.

SFEs of LSBGs are lower than 1.364 $\times$ 10$^{-10}$ yr$^{-1}$ and
are far lower than 5.25 $\times$ 10$^{-10}$ yr$^{-1}$ observed in
normal spiral galaxies \citep{2008AJ....136.2782L}. Generally,
molecular gas is directly related to star formation, so SFE can
reflect the efficiency of transforming hydrogen atom to molecular
hydrogen. Low SFE may hint that atomic hydrogen produces molecular
hydrogen at a low speed. In short, the low SFR and SFE imply a lack
of molecular hydrogen gas in LSBGs and this will be discussed in \textsection 4.

\section{Discussion \label{dis}}

\subsection{Comment of Individual Galaxies \label{CIG}}

\textbf{AGCNr12289:} is a late-type spiral galaxy with a optical disk size of about 35$''$
and has a potential AGN/LINER core
\citep{1998AJ....116.1650S}. There is a supernova (SN2002en) in
AGCNr12289. \citet{2003ApJ...588..230O} has detected
CO(J=1-0,J=2-1) content in AGCNr12289 using the IRAM 30-meter
telescope.
The M$_{H_2}$ of AGCNr12289 is about 15.8 $\times$ 10$^7\,$M$_\odot$ using
CO(J=2-1) observed by IRAM, which is lower than our upper limit
of 41.2 $\times$ 10$^7\,$M$_\odot$. The ratio of M$_{H_2}$/M$_{HI}$ is 0.008 for AGCNr12289.

\textbf{AGCNr188845},\textbf{AGCNr102243},\textbf{AGCNr102635},\textbf{AGCNr102981}:
For these galaxies, the optical sizes are about 54$''$, 52$''$, 46$''$ and 46$''$. From our calculations, the M$_{H_2}$/M$_{HI}$ are lower
than 0.044, 0.017, 0.048 and 0.018, respectively. The metallicity of AGCNr188845 is 0.01 and 1/2 of solar metallicity. AGCNr102981
is a typical late-type spiral galaxy and the spiral structures are
shown in Figure 1. 

\textbf{AGCNr4528}, \textbf{AGCNr12212}, \textbf{AGCNr110150}: The
optical sizes of these three galaxies are about 66$''$, 62$''$ and 80$''$ respectively which indicate that JCMT's beam size is not
enough to cover their total CO content. The metallicity of AGCNr4528 is 0.02 and same to solar metallicity. The M$_{H_2}$/M$_{HI}$ for these three LSBGs are lower than 0.018, 0.0201 and 0.022 respectively.

\textbf{NGC7589}: The optical size of NGC 7589 is 76$''$. It
is a Seyfert-1 galaxy. The SFR from NUV band is 1.003$\,$M$_\odot$yr$^{-1}$ and its metallicity is 0.04 and twice the solar metallicity.
There are some works investigating this galaxy. It is a typical
giant LSBG and has type 1 XUV-disk galaxy\citep{2008ApJ...681..244B}.
The upper limit of M$_{H2}$ is 8.25 $\times$ 10$^8\,$M$_\odot$ and the ratio of M$_{H_2}$/M$_{HI}$ is lower
than 0.081.

\subsection{M$_{H_2}$ Versus L$_{12\mu m}$ \label{MVL}}
The 12$\,\mu m$ (WISE W3) band is an effective probe of star
formation \citep{2012ApJ...748...80D} and has a good linear
relationship with molecular hydrogen mass for SF galaxies
\citep{2015ApJ...799...92J}. In this section, we try to explore the
relation between molecular gas and 12$\,\mu$m emission in LSBGs.

\begin{figure}[h!tb]
\captionstyle{flushleft}
\onelinecaptionstrue
\begin{center}
\includegraphics[width=8.5in]{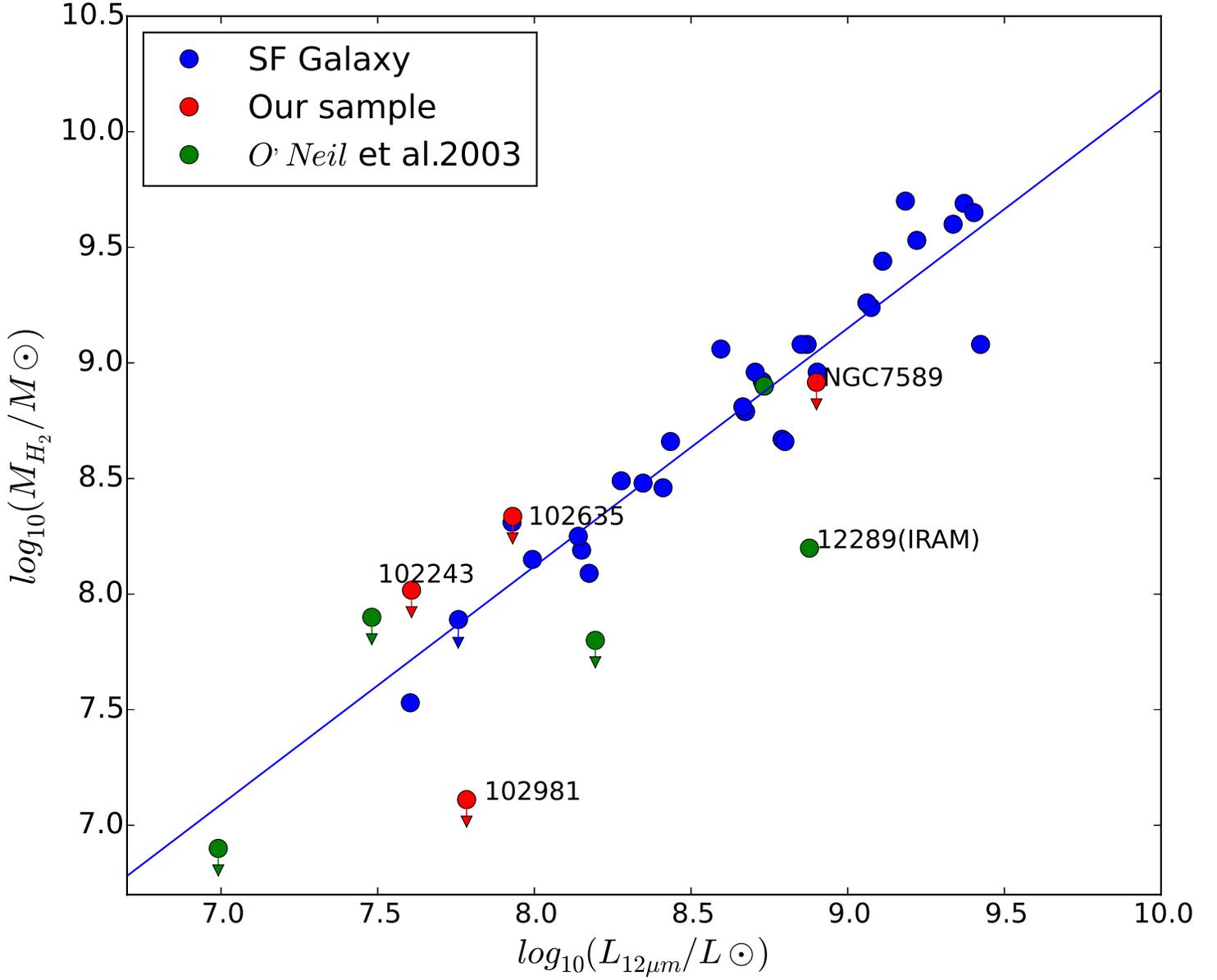}
\end{center}
\captionsetup{margin=80pt} \caption{Luminosity of 12$\,\mu$m vs.
M$_{H_2}$. The red dots
are LSBGs from this work, green dots are from
\citet{2003ApJ...588..230O} and the blue dots are SF galaxies from \citet{2015ApJ...799...92J}. The arrows indicate the upper limits.
       \label{MH2}}
\end{figure}

Five sources in our sample have 12$\,\mu$m emission data provided by
the WISE ALL-SKY Survey. 
Figure \ref{MH2} shows L$_{12\mu m}$ versus M$_{H_2}$. The blue line in Figure \ref{MH2} is the relationship between M$_{H_2}$ (CO$_{J=2-1}$) and luminosity of 12$\,\mu$m for SF galaxies\citep{2015ApJ...799...92J}.

AGCNr12289 and NGC7589 have higher luminosities of 12$\,\mu$m in
Figure \ref{MH2}. AGNCr12289 has a potential AGN/LINER core and NGC
7589 is a Seyfert 1 galaxy which may cause them to be different from
other LSBGs in our sample. The LSBGs show low luminosities of 12$\,\mu$m and it may mean less dust in LSBGs. For SF galaxies,
they always have larger 12$\,\mu$m luminosities compared with LSBGs.

\subsection{The Stellar Mass and Gas Content\label{MXG}}
Six LSBGs of our sample have 3.4$\,\mu m$ (WISE W1) band data.
Using the method in \citet{2013MNRAS.433.2946W} by following
Equation (5), we adopt 3.4$\,\mu$m data to calculate the stellar
mass:

log$_{10}$(M$_*$/M$_\odot$)=(-0.040$\pm$0.001) + (1.120$\pm$0.001) $\times$ log$_{10}$($\nu$ L$_{\nu}$(3.4$\mu$m)/L$_\odot$) $\quad$(5)

As the redshifts of our sample are small, we ignore the
k-correction. Their stellar masses are about (1.41-83.17)
$\times$ 10$^8$M$_\odot$. LSBGs usually have low masses. The stellar
masses of our sample are shown in Table \ref{tbl3}.

The stellar masses of SF galaxies have a different relation with
molecular gas and atomic gas. The ratios of M$_{H_2}$/M$_*$ are almost
constant \citep{2015ApJ...799...92J}. We are more interested in the
relation between stellar mass and gas content of LSBGs.

Figure \ref{stellarmass} shows the relation between stellar mass
M$_*$ and gas content in SF galaxies and our LSBG sample. The M$_{H_2}$/M$_*$ shows a flat trend within the considerable scatter and M$_{HI}$/M$_*$ shows an obvious
decline with increasing stellar mass. It seems that the molecular
gas fraction (M$_{H_2}$/M$_*$) is similar in LSBGs and in SF galaxies,
although our results just provide upper limits except for the AGCNr12289.
Compared with SF galaxies, the stellar mass of our sample is mainly lower than
10$^{9}\,$M$_\odot$ excluding the two special galaxies (AGCNr12289 and
NGC7589). Stellar masses of AGCNr12289 and NGC7589 may be
overestimated by the central AGN.

\subsection{M$_{H_2}$/M$_{HI}$ and SFR/M$_{H_2}$\label{MXG}}
HI gas is the original material involved in star formation, and star
formation is directly related to molecular gas. The
M$_{H_2}$/M$_{HI}$ ratio should be different for different galaxies. The gas content could change along the Main-Sequence of star-forming galaixes. Other factors,such as metallicity, environment could affect the ratio. Even for the same galaxy, its gas fraction would also change during its different evolutionary phase.
In our nine LSBGs, the M$_{H_2}$/M$_{HI}$ ratios are less than 0.02.
In typically brighter Sd-Sm spirals \citep{1989ApJ...347L..55Y}, the
M$_{H_2}$/M$_{HI}$ ratio is about 0.2.

In the section 3.1, we have discussed the M$_{HI}$ and M$_{H_2}$ in
SF galaxies and LSBGs. The M$_{H_2}$ in LSBGs is lower than that in
SF galaxies. In this section we compare M$_{H_2}$/M$_{HI}$ and
SFR/M$_{H_2}$ in those LSBGs and SF galaxies.

Figure \ref{counts} show the distribution of M$_{H_2}$/M$_{HI}$ on the left and SFR/M$_{H_2}$ on the right. In right panel, NGC7589 has completely different values of SFR/M$_{H_2}$ from the other LSBGs because of AGN influence.
Due to the dispersed distribution,
the difference in SFR/M$_{H_2}$ between SF galaxies and LSBGs
is not quite obvious.

We can see that the ratios of M$_{H_2}$/M$_{HI}$ in LSBGs are less
than those in SF galaxies. The rate of transforming atomic hydrogen
to molecular hydrogen in LSBGs is lower than that in SF galaxies. In
Section 3.1, we also have discussed the shortage of molecular gas in the
special environment of LSBGs. However, CO is more easily
photodissociated than H$_2$ in a metal-poor environment
\citep{2010ApJ...716.1191W,2011MNRAS.412.1686S}. Thus, the conversion factor of CO-to-H$_2$ is usually higher than that in SF galaxies.
In this work, we adopt the factor 3.162 $\times$ 10$^{20}\,$cm$^{-2}$/(K$\,$kms$^{-1})$
to be consistent with previous works. According to the work of \citet{2012MNRAS.421.3127N},
the factor may be larger than 3.16 $\times$ 10$^{20}\,$cm$^{-2}$/(K$\,$kms$^{-1})$ and up to 15 $\times$ 10$^{20}\,$cm$^{-2}$/(K$\,$kms$^{-1})$. So, the M$_{H_2}$ of our sample may be underestimated.
Although lacking a detected CO content, it is still possible that M$_{H_2}$ is underestimated in LSBGs.

\section{Summary \label{sum}}
We observed CO(J=2-1) emission lines in nine LSBGs with JCMT and
H$\alpha$ images with the 2.16-meter telescope administered by NAOC. As no CO has been detected, only upper limits on the H$_2$ masses are given. The upper limits of hydrogen molecular masses are about (1.2-82.4) $\times$ 10$^7\,$M$_\odot$.
Their star formation rates are about 0.056-0.83$\,$M$_\odot$yr$^{-1}$ and 0.146-1.003$\,$M$_\odot$yr$^{-1}$ estimated by H$\alpha$ and NUV luminosities, respectively. The steller masses are about (0.14-8.31) $\times$ 10$^9\,$M$_\odot$ estimated by the WISE 3.4$\,\mu$m band.
From our results, the M$_{H_2}$ and stellar mass in LSBGs are lower than those in SF galaxies. Low SFRs in LSBGs may be related to low molecular hydrogen mass,
which may indicate low productivity of atomic hydrogen transforming
into molecular hydrogen. More direct detection of molecular gas of LSBG is the key to answer the question in the future.

\section*{Acknowledgments}
We would like to thank the referee for helpful suggestions.
We thank the Key Laboratory of
Optical Astronomy and Xinglong Observing Station administered by
NAOC for their help during observations. We also thank the ALFALFA
team for providing the $\alpha$.40 catalog, and the WISE team and
GALEX team for their wonderful released data. We thank Dr.Xianzhong Zheng for given suggestions.

This project is supported by the National Natural Science Foundation
of China (Grant Nos. 11403037, 11173030, 11225316, 11503013,
11403061, U1531245, 11303038); This project is also supported by the
Strategic Priority Research Program, "The Emergence of Cosmological
Structures" of the Chinese Academy of Sciences (Grant
No.XDB09000000) and the China Ministry of Science and Technology under the State Key Development
Program for Basic Research (Grant Nos. 2012CB821803, 2014CB845705).
This work is also sponsored in part by the Chinese Academy of Sciences (CAS), through a grant to the CAS South America Center for Astronomy (CASSACA) in Santiago, Chile.

The James Clerk Maxwell Telescope is operated by the East Asian Observatory on behalf of the National Astronomical Observatories of China and the Chinese Academy of Sciences, the National Astronomical Observatory of Japan, Academia Sinica Institute of Astronomy and Astrophysics, the Korea Astronomy and Space Science Institute, with additional funding support from the Science and Technology Facilities Council of the United Kingdom and participating universities in the United Kingdom and Canada. Our JCMT project ID is M15BI057

\clearpage

\bibliographystyle{apj}
\bibstyle{thesisstyle}
\bibliography{main.bib}

\begin{thebibliography}{56}
\expandafter\ifx\csname natexlab\endcsname\relax\def\natexlab#1{#1}\fi

\bibitem[{{Blanton} \& {Moustakas}(2009)}]{2009ARA&A..47..159B}
{Blanton}, M.~R. \& {Moustakas}, J. 2009, \araa, 47, 159

\bibitem[{{Boissier} {et~al.}(2008){Boissier}, {Gil de Paz}, \&
  {Boselli}}]{2008ApJ...681..244B}
{Boissier}, S., {Gil de Paz}, A., \& {Boselli}, A. 2008, \apj, 681, 244

\bibitem[{{Bolatto} {et~al.}(2013){Bolatto}, {Wolfire}, \&
  {Leroy}}]{2013ARA&A..51..207B}
{Bolatto}, A.~D., {Wolfire}, M., \& {Leroy}, A.~K. 2013, \araa, 51, 207

\bibitem[{{Braine} {et~al.}(2000){Braine}, {Herpin}, \&
  {Radford}}]{2000A&A...358..494B}
{Braine}, J., {Herpin}, F., \& {Radford}, S.~J.~E. 2000, \aap, 358, 494

\bibitem[{{Burkholder} {et~al.}(2001){Burkholder}, {Impey}, \&
  {Sprayberry}}]{2001AJ....122.2318B}
{Burkholder}, V., {Impey}, C., \& {Sprayberry}, D. 2001, \aj, 122, 2318

\bibitem[{{Das} {et~al.}(2010){Das}, {Boone}, \&
  {Viallefond}}]{2010A&A...523A..63D}
{Das}, M., {Boone}, F., \& {Viallefond}, F. 2010, \aap, 523, A63

\bibitem[{{Das} {et~al.}(2007){Das}, {Kantharia}, {Ramya}, {Prabhu}, {McGaugh},
  \& {Vogel}}]{2007MNRAS.379...11D}
{Das}, M., {Kantharia}, N., {Ramya}, S., {Prabhu}, T.~P., {McGaugh}, S.~S., \&
  {Vogel}, S.~N. 2007, \mnras, 379, 11

\bibitem[{{de Blok} {et~al.}(1996){de Blok}, {McGaugh}, \& {van der
  Hulst}}]{1996MNRAS.283...18D}
{de Blok}, W.~J.~G., {McGaugh}, S.~S., \& {van der Hulst}, J.~M. 1996, \mnras,
  283, 18

\bibitem[{{de Blok} \& {van der
  Hulst}(1998{\natexlab{a}})}]{1998A&A...335..421D}
{de Blok}, W.~J.~G. \& {van der Hulst}, J.~M. 1998{\natexlab{a}}, \aap, 335,
  421

\bibitem[{{de Blok} \& {van der
  Hulst}(1998{\natexlab{b}})}]{1998A&A...336...49D}
---. 1998{\natexlab{b}}, \aap, 336, 49

\bibitem[{{de Blok} {et~al.}(1995){de Blok}, {van der Hulst}, \&
  {Bothun}}]{1995MNRAS.274..235D}
{de Blok}, W.~J.~G., {van der Hulst}, J.~M., \& {Bothun}, G.~D. 1995, \mnras,
  274, 235

\bibitem[{{de Vaucouleurs} {et~al.}(1991){de Vaucouleurs}, {de Vaucouleurs},
  {Corwin}, {Buta}, {Paturel}, \& {Fouqu{\'e}}}]{1991rc3..book.....D}
{de Vaucouleurs}, G., {de Vaucouleurs}, A., {Corwin}, Jr., H.~G., {Buta},
  R.~J., {Paturel}, G., \& {Fouqu{\'e}}, P. 1991, {Third Reference Catalogue of
  Bright Galaxies. Volume I: Explanations and references. Volume II: Data for
  galaxies between 0$^{h}$ and 12$^{h}$. Volume III: Data for galaxies between
  12$^{h}$ and 24$^{h}$.}

\bibitem[{{Donoso} {et~al.}(2012){Donoso}, {Yan}, {Tsai}, {Eisenhardt},
  {Stern}, {Assef}, {Leisawitz}, {Jarrett}, \&
  {Stanford}}]{2012ApJ...748...80D}
{Donoso}, E., {Yan}, L., {Tsai}, C., {Eisenhardt}, P., {Stern}, D., {Assef},
  R.~J., {Leisawitz}, D., {Jarrett}, T.~H., \& {Stanford}, S.~A. 2012, \apj,
  748, 80

\bibitem[{{Du} {et~al.}(2015){Du}, {Wu}, {Lam}, {Zhu}, {Lei}, \&
  {Zhou}}]{2015AJ....149..199D}
{Du}, W., {Wu}, H., {Lam}, M.~I., {Zhu}, Y., {Lei}, F., \& {Zhou}, Z. 2015,
  \aj, 149, 199

\bibitem[{{Epinat} {et~al.}(2008){Epinat}, {Amram}, \&
  {Marcelin}}]{2008MNRAS.390..466E}
{Epinat}, B., {Amram}, P., \& {Marcelin}, M. 2008, \mnras, 390, 466

\bibitem[{{Fan} {et~al.}(2016){Fan}, {Wang}, \& {Jiang}}]{2016arXiv160509166F}
{Fan}, Z., {Wang}, H., \& {Jiang}, X. 2016, ArXiv e-prints

\bibitem[{{Gerritsen} \& {de Blok}(1999)}]{1999A&A...342..655G}
{Gerritsen}, J.~P.~E. \& {de Blok}, W.~J.~G. 1999, \aap, 342, 655

\bibitem[{{Giovanelli} {et~al.}(2005){Giovanelli}, {Haynes}, \&
  {Kent}}]{2005AJ....130.2598G}
{Giovanelli}, R., {Haynes}, M.~P., \& {Kent}, B.~R. 2005, \aj, 130, 2598

\bibitem[{{Haynes} {et~al.}(2011){Haynes}, {Giovanelli}, \&
  {Martin}}]{2011AJ....142..170H}
{Haynes}, M.~P., {Giovanelli}, R., \& {Martin}, A.~M. 2011, \aj, 142, 170

\bibitem[{{Hirst} \& {Cavanagh}(2005)}]{2005StaUN.236.....H}
{Hirst}, P. \& {Cavanagh}, B. 2005, Starlink User Note, 236

\bibitem[{{Impey} \& {Bothun}(1997)}]{1997ARA&A..35..267I}
{Impey}, C. \& {Bothun}, G. 1997, \araa, 35, 267

\bibitem[{{Jiang} {et~al.}(2015){Jiang}, {Wang}, {Gu}, {Wang}, \&
  {Zhang}}]{2015ApJ...799...92J}
{Jiang}, X.-J., {Wang}, Z., {Gu}, Q., {Wang}, J., \& {Zhang}, Z.-Y. 2015, \apj,
  799, 92

\bibitem[{{Jimenez} {et~al.}(1998){Jimenez}, {Padoan}, {Matteucci}, \&
  {Heavens}}]{1998MNRAS.299..123J}
{Jimenez}, R., {Padoan}, P., {Matteucci}, F., \& {Heavens}, A.~F. 1998, \mnras,
  299, 123

\bibitem[{{Kennicutt}(1998)}]{1998ARA&A..36..189K}
{Kennicutt}, Jr., R.~C. 1998, \araa, 36, 189

\bibitem[{{Knezek}(1993)}]{1993PhDT.........2K}
{Knezek}, P.~M. 1993, PhD thesis, Massachusetts Univ., Amherst.

\bibitem[{{Lauberts} \& {Valentijn}(1989)}]{1989spce.book.....L}
{Lauberts}, A. \& {Valentijn}, E.~A. 1989, {The surface photometry catalogue of
  the ESO-Uppsala galaxies}

\bibitem[{{Leroy} {et~al.}(2009){Leroy}, {Walter}, \&
  {Bigiel}}]{2009AJ....137.4670L}
{Leroy}, A.~K., {Walter}, F., \& {Bigiel}, F. 2009, \aj, 137, 4670

\bibitem[{{Leroy} {et~al.}(2008){Leroy}, {Walter}, \&
  {Brinks}}]{2008AJ....136.2782L}
{Leroy}, A.~K., {Walter}, F., \& {Brinks}, E. 2008, \aj, 136, 2782

\bibitem[{{Matthews} \& {Gao}(2001)}]{2001ApJ...549L.191M}
{Matthews}, L.~D. \& {Gao}, Y. 2001, \apjl, 549, L191

\bibitem[{{Matthews} {et~al.}(2005){Matthews}, {Gao}, {Uson}, \&
  {Combes}}]{2005AJ....129.1849M}
{Matthews}, L.~D., {Gao}, Y., {Uson}, J.~M., \& {Combes}, F. 2005, \aj, 129,
  1849

\bibitem[{{Matthews} {et~al.}(2001){Matthews}, {van Driel}, \&
  {Monnier-Ragaigne}}]{2001A&A...365....1M}
{Matthews}, L.~D., {van Driel}, W., \& {Monnier-Ragaigne}, D. 2001, \aap, 365,
  1

\bibitem[{{Matthews} \& {Wood}(2001)}]{2001ApJ...548..150M}
{Matthews}, L.~D. \& {Wood}, K. 2001, \apj, 548, 150

\bibitem[{{McGaugh}(1994)}]{1994ApJ...426..135M}
{McGaugh}, S.~S. 1994, \apj, 426, 135

\bibitem[{{McGaugh} \& {Bothun}(1994)}]{1994AJ....107..530M}
{McGaugh}, S.~S. \& {Bothun}, G.~D. 1994, \aj, 107, 530

\bibitem[{{McGaugh} \& {de Blok}(1997)}]{1997ApJ...481..689M}
{McGaugh}, S.~S. \& {de Blok}, W.~J.~G. 1997, \apj, 481, 689

\bibitem[{{Narayanan} {et~al.}(2012){Narayanan}, {Krumholz}, {Ostriker}, \&
  {Hernquist}}]{2012MNRAS.421.3127N}
{Narayanan}, D., {Krumholz}, M.~R., {Ostriker}, E.~C., \& {Hernquist}, L. 2012,
  \mnras, 421, 3127

\bibitem[{{O'Neil} {et~al.}(2004){O'Neil}, {Bothun}, {van Driel}, \& {Monnier
  Ragaigne}}]{2004A&A...428..823O}
{O'Neil}, K., {Bothun}, G., {van Driel}, W., \& {Monnier Ragaigne}, D. 2004,
  \aap, 428, 823

\bibitem[{{O'Neil} {et~al.}(2000){O'Neil}, {Hofner}, \&
  {Schinnerer}}]{2000ApJ...545L..99O}
{O'Neil}, K., {Hofner}, P., \& {Schinnerer}, E. 2000, \apjl, 545, L99

\bibitem[{{O'Neil} {et~al.}(2003){O'Neil}, {Schinnerer}, \&
  {Hofner}}]{2003ApJ...588..230O}
{O'Neil}, K., {Schinnerer}, E., \& {Hofner}, P. 2003, \apj, 588, 230

\bibitem[{{Peebles}(2001)}]{2001ApJ...557..495P}
{Peebles}, P.~J.~E. 2001, \apj, 557, 495

\bibitem[{{Pickering} {et~al.}(1997){Pickering}, {Impey}, {van Gorkom}, \&
  {Bothun}}]{1997AJ....114.1858P}
{Pickering}, T.~E., {Impey}, C.~D., {van Gorkom}, J.~H., \& {Bothun}, G.~D.
  1997, \aj, 114, 1858

\bibitem[{{Pustilnik} {et~al.}(2011){Pustilnik}, {Martin}, {Tepliakova}, \&
  {Kniazev}}]{2011MNRAS.417.1335P}
{Pustilnik}, S.~A., {Martin}, J.-M., {Tepliakova}, A.~L., \& {Kniazev}, A.~Y.
  2011, \mnras, 417, 1335

\bibitem[{{Savage} \& {Mathis}(1979)}]{1979ARA&A..17...73S}
{Savage}, B.~D. \& {Mathis}, J.~S. 1979, \araa, 17, 73

\bibitem[{{Schombert}(1998)}]{1998AJ....116.1650S}
{Schombert}, J. 1998, \aj, 116, 1650

\bibitem[{{Schombert} {et~al.}(1990){Schombert}, {Bothun}, {Impey}, \&
  {Mundy}}]{1990AJ....100.1523S}
{Schombert}, J.~M., {Bothun}, G.~D., {Impey}, C.~D., \& {Mundy}, L.~G. 1990,
  \aj, 100, 1523

\bibitem[{{Schruba} {et~al.}(2012){Schruba}, {Leroy}, \&
  {Walter}}]{2012AJ....143..138S}
{Schruba}, A., {Leroy}, A.~K., \& {Walter}, F. 2012, \aj, 143, 138

\bibitem[{{Shetty} {et~al.}(2011){Shetty}, {Glover}, {Dullemond}, \&
  {Klessen}}]{2011MNRAS.412.1686S}
{Shetty}, R., {Glover}, S.~C., {Dullemond}, C.~P., \& {Klessen}, R.~S. 2011,
  \mnras, 412, 1686

\bibitem[{{Subramanian} {et~al.}(2016){Subramanian}, {Ramya}, {Das}, {George},
  {Sivarani}, \& {Prabhu}}]{2016MNRAS.455.3148S}
{Subramanian}, S., {Ramya}, S., {Das}, M., {George}, K., {Sivarani}, T., \&
  {Prabhu}, T.~P. 2016, \mnras, 455, 3148

\bibitem[{{van der Hulst} {et~al.}(1993){van der Hulst}, {Skillman}, {Smith},
  {Bothun}, {McGaugh}, \& {de Blok}}]{1993AJ....106..548V}
{van der Hulst}, J.~M., {Skillman}, E.~D., {Smith}, T.~R., {Bothun}, G.~D.,
  {McGaugh}, S.~S., \& {de Blok}, W.~J.~G. 1993, \aj, 106, 548

\bibitem[{{van Zee}(2000)}]{2000AJ....119.2757V}
{van Zee}, L. 2000, \aj, 119, 2757

\bibitem[{{Wen} {et~al.}(2013){Wen}, {Wu}, \& {Zhu}}]{2013MNRAS.433.2946W}
{Wen}, X.-Q., {Wu}, H., \& {Zhu}, Y.-N. 2013, \mnras, 433, 2946

\bibitem[{{Wolfire} {et~al.}(2010){Wolfire}, {Hollenbach}, \&
  {McKee}}]{2010ApJ...716.1191W}
{Wolfire}, M.~G., {Hollenbach}, D., \& {McKee}, C.~F. 2010, \apj, 716, 1191

\bibitem[{{Wu} {et~al.}(2002){Wu}, {Burstein}, \& {Deng}}]{2002AJ....123.1364W}
{Wu}, H., {Burstein}, D., \& {Deng}, Z. 2002, \aj, 123, 1364

\bibitem[{{Wyder} {et~al.}(2009){Wyder}, {Martin}, \&
  {Barlow}}]{2009ApJ...696.1834W}
{Wyder}, T.~K., {Martin}, D.~C., \& {Barlow}, T.~A. 2009, \apj, 696, 1834

\bibitem[{{Young} \& {Knezek}(1989)}]{1989ApJ...347L..55Y}
{Young}, J.~S. \& {Knezek}, P.~M. 1989, \apjl, 347, L55

\bibitem[{{Zheng} {et~al.}(1999){Zheng}, {Shang}, \&
  {Su}}]{1999AJ....117.2757Z}
{Zheng}, Z., {Shang}, Z., \& {Su}, H. 1999, \aj, 117, 2757

\end{thebibliography}

\clearpage
\newgeometry{left=0.1cm,bottom=1cm}

\clearpage
\newgeometry{left=0.0cm,bottom=1cm}

\clearpage
\newgeometry{left=0.0cm,bottom=1cm}

\clearpage
\newgeometry{left=0.0cm,bottom=1cm}

\clearpage
\newgeometry{left=0.0cm,bottom=1cm}
\begin{figure}[h!tb]
\captionstyle{flushleft}
\onelinecaptionstrue
\begin{center}
\includegraphics[width=8.3in]{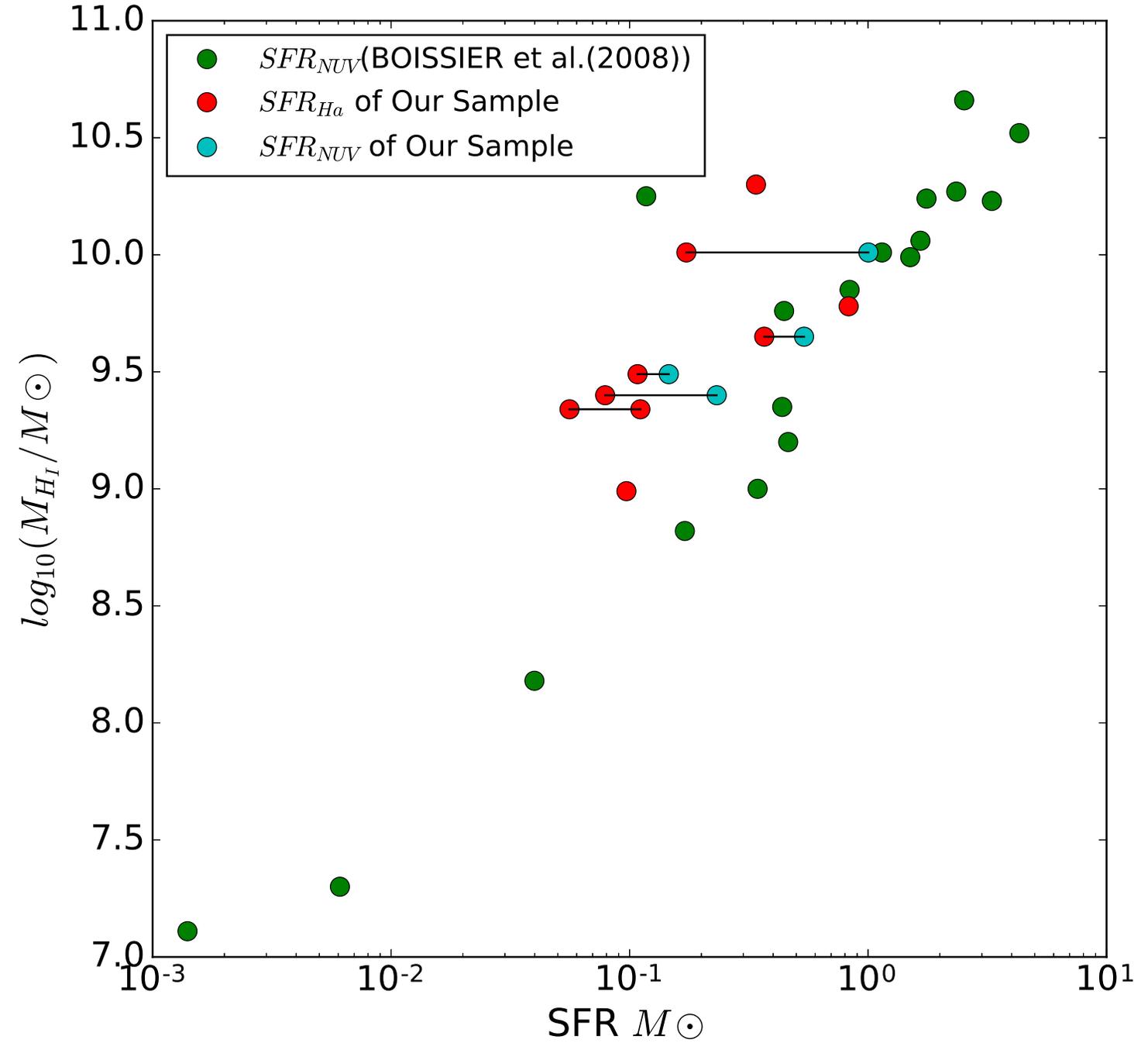}
\end{center}
\captionsetup{margin=80pt} \caption{ HI mass vs. SFR. The red dots
are SFR calculated by H$\alpha$ and the blue dots are SFR calculated
by NUV luminosity. The black line connects the same target.
      The green dots are the sample of LSBGs from \citet{2008ApJ...681..244B}.
       \label{SFR}}

\end{figure}

\clearpage
\newgeometry{left=0.0cm,bottom=1cm}
\begin{figure}[h!tb]
\captionstyle{flushleft}
\onelinecaptionstrue
\begin{center}
\includegraphics[width=8.5in]{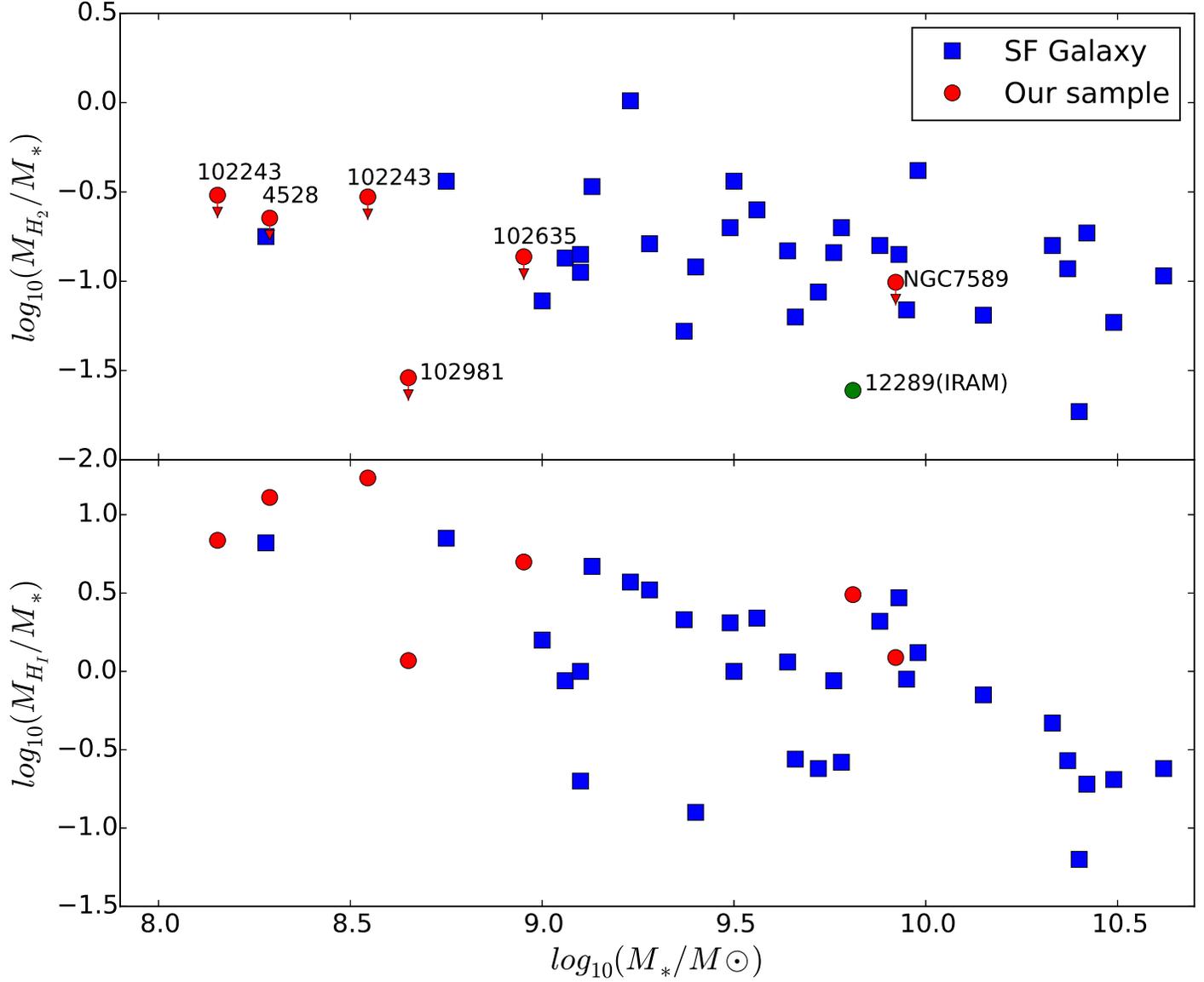}
\end{center}
\captionsetup{margin=80pt} \caption{ Stellar mass (M$_*$) vs. the
ratio of M$_{H_2}$/M$_*$ (upper) and M$_{HI}$/M$_*$ (lower). The red
dots are LSBGs from this work and the blue squares are SF
galaxies\citep{2015ApJ...799...92J}. The arrows indicate the upper limits.
       \label{stellarmass}}

\end{figure}

\clearpage
\newgeometry{left=0.0cm,bottom=1cm}
\begin{figure}[h!tb]
\captionstyle{flushleft}
\onelinecaptionstrue
\begin{center}
\includegraphics[width=8.5in]{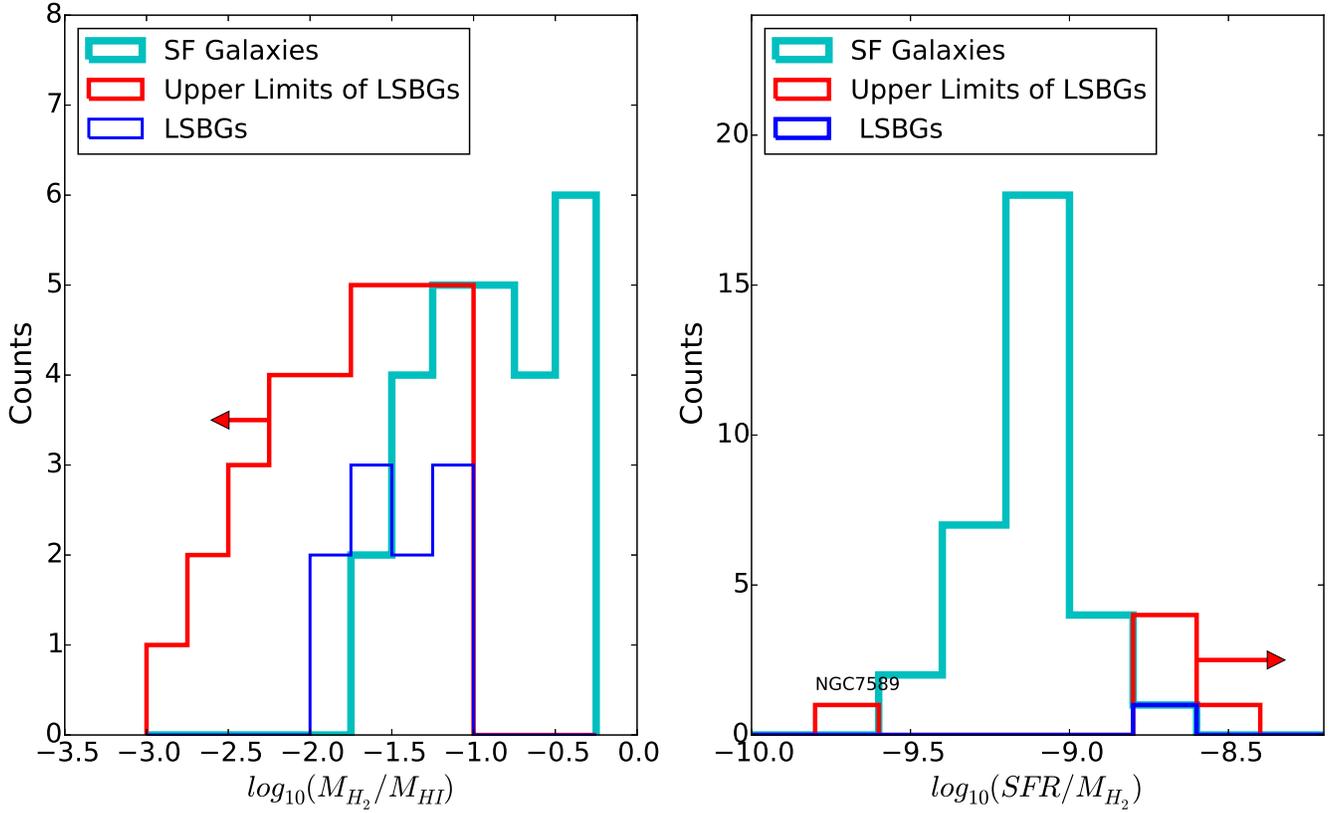}
\end{center}
\captionsetup{margin=80pt} \caption{Left panel: The distribution of
 M$_{H_2}$/M$_{HI}$. Right panel: The distribution of SFR/M$_{H_2}$.
 The red histogram represents upper limits of LSBGs and the blue histogram represents those detected molecular LSBGs. All LSBGs data are from our sample and previous works\citep{1990AJ....100.1523S,1998A&A...335..421D,2000ApJ...545L..99O,2000A&A...358..494B,2003ApJ...588..230O}. The cyan histogram represents SF galaxies\citep{2015ApJ...799...92J}. The arrows means upper limits.
       \label{counts}}

\end{figure}

\clearpage
%\LongTables
\begin{deluxetable}{ccccccccccccccccc}
\tabletypesize{\small} \rotate \tablecolumns{15}
\tablecaption{Parameters of Low Surface brightness Galaxies
\label{tbl1}} \tablewidth{0pt} \tablehead{
\colhead{Name} &\colhead{RA(J2000)} & \colhead{Dec(J2000)} & \colhead{V$_{HeI}$ }  & \colhead{$u_{B(0)obs}$ } & \colhead{{W$_{50}$}\tablenotemark{a}} & \colhead{logM$_{HI}$} & \colhead{D} &\colhead{Major-axis} & \colhead{z} & \colhead{NUV$_{AB}$} &\colhead{r$_{SDSS}$}& \colhead{{W1}\tablenotemark{b}}& \colhead{{W3}\tablenotemark{c}}  \\
  &  &  & \colhead{kms$^{-1}$}  & \colhead{mag$\,$arcsec$^{-2}$} & \colhead{kms$^{-1}$} & \colhead{M$_\odot$} & \colhead{Mpc} &\colhead{arcsec} &  &\colhead{mag} &\colhead{mag} &\colhead{mag} &\colhead{mag}\\
}

\startdata
AGCNr188845 & 08:01:13.02  & +11:37:24.48 & 4935 & 23.056 & 79 & 8.99 & 72.5 & 54 &  0.0164 &              &16.30 & 13.923     & \\
            &              &              &      &        &    &       &      &       &          &             &$\pm$0.02 & $\pm$0.044 & \\
AGCNr4528   & 08:40:58.60  & +16:11:00.14 & 4288 & 22.900 & 88 & 9.40 & 63.6 & 66 &  0.0143 &17.56         &15.90 & 13.335     & \\
            &              &              &      &        &    &       &      &       &          &$\pm$0.02    &$\pm$0.01 & $\pm$0.028 & \\
AGCNr12212  & 22:50:30.26  & +29:08:19.52 &  894 & 23.040 & 99 & 9.34 & 24.3 & 62 &  0.0029 &              &15.02 &            & \\
            &              &              &      &        &    &       &      &       &          &             &$\pm$0.00 &            & \\
AGCNr110150 & 01:14:45.56  & +27:08:11.10 & 3617 & 22.760 &108 & 9.49 & 49.5 & 80  &  0.0120 &17.52         &16.01 &            & \\
            &              &              &      &        &    &       &      &       &          &$\pm$0.03    &$\pm$0.01 &            & \\
AGCNr102243 & 00:05:05.06  & +23:58:14.09 & 6575 & 22.500 &139 & 9.78 & 89.0 & 52 &  0.0219 &              &20.59 & 13.494     &10.83\\
            &              &              &      &        &    &       &      &       &          &             &$\pm$0.04 & $\pm$0.019 &$\pm$0.336 \\
AGCNr12289  & 22:59:41.52  & +24:04:29.77 &10165 & 22.650 &217 & 10.3 &140.2 & 35 &  0.0339 &              &14.60 & 11.658     & 8.64\\
            &              &              &      &        &    &       &      &       &          &             &$\pm$0.00 & $\pm$0.007 &$\pm$0.034 \\
AGCNr102981 & 00:02:55.56  & +28:16:38.78 & 4583 & 22.540 & 65 & 8.72 & 66.2 &46   &  0.0152 &              &15.41 & 12.615     & 9.74\\
            &              &              &      &        &    &       &      &       &          &             &$\pm$0.00 & $\pm$0.012 &$\pm$0.117 \\
AGCNr102635 & 00:16:12.22  & +24:50:59.04 & 9491 & 22.573 & 94 & 9.65 &138.7 & 46  &  0.0316 &18.34         &16.66 & 13.551     & 10.98\\
            &              &              &      &        &    &       &      &       &          &$\pm$0.04    &$\pm$0.01 & $\pm$0.026 &$\pm$0.261 \\
NGC7589     & 23:18:15.60  & +00:15:40.00 & 8938 & 25.000*\tablenotemark{d} &345 &10.01 &120.7 & 76  &  0.0298 &17.53         &14.08 & 11.084     & 8.26\\
            &              &              &      &        &    &       &      &       &          &$\pm$0.01    &$\pm$0.00 & $\pm$0.009 &$\pm$0.024 \\

\enddata
\tablenotetext{a}{${W_5}_0$ is the Full Width at Half Maxinum(FWHM) of HI emission line.}
\tablenotetext{b}{W1 is 3.4$\mu$m band of the WISE.}
\tablenotetext{c}{W3 is 12$\mu$m band of the WISE.}
\tablenotetext{d}{For NGC 7589, its surface brightness is from \citet{1991rc3..book.....D}.}
\end{deluxetable}
\clearpage

\begin{deluxetable}{cccccccccc}
\tabletypesize{\small}
\rotate
\tablecolumns{10}
\tablecaption{Observation Details \label{tbl2}}
\tablewidth{0pt}
\tablehead{
\colhead{Name} &\colhead{Receiver} & \colhead{Weather band} &\colhead{Frequency} & \colhead{Integration time} & \colhead{rms noise } & \colhead{Facility} & \colhead{{H$\alpha$ filter}\tablenotemark{a}} & \colhead{Exp$_{H\alpha}$} & \colhead{Exp$_{R_{band}}$} \\
 & & &\colhead{GHZ} &\colhead{hour} & \colhead{mk}& &$\AA$  & \colhead{s}&  \colhead{s}\\
}
\startdata
AGCNr188845 & RxA3/JCMT &4 &226.91 &1.63 & 2.7 & BFOSC/2.16 &6660.0 & 1800 & 600\\
AGCNr4528   & RxA3/JCMT &4 &227.29 &2.62 & 2.0 & BFOSC/2.16 &6660.0 & 1800 & 600\\
AGCNr12212  & RxA3/JCMT &5 &229.86 &7.8  & 2.1 &       &            &      &\\
AGCNr110150 & RxA3/JCMT &5 &227.67 &7.4  & 2.8 & BFOSC/2.16 &6660.0 & 1800 & 600\\
AGCNr102243 & RxA3/JCMT &5 &225.60 &7.8  & 2.6 & BFOSC/2.16 &6710.0 & 1800 & 600\\
AGCNr12289  & RxA3/JCMT &5 &222.99 &7.6  & 3.3 & BFOSC/2.16 &6760.0 & 1800 & 600\\
AGCNr102981 & RxA3/JCMT &5 &227.06 &7.8  & 2.0 & BFOSC/2.16 &6660.0 & 1800 & 600\\
AGCNr102635 & RxA3/JCMT &5 &223.47 &8.3  & 3.1 & BFOSC/2.16 &6760.0 & 1800 & 600\\
NGC7589     & RxA3/JCMT &5 &223.87 &7.8  & 2.1 &  &   & &\\
\enddata
\tablenotetext{a}{The central wavelength of H$\alpha$ filter.}
\end{deluxetable}
\clearpage

\begin{deluxetable}{ccccccccccc}
\tabletypesize{\footnotesize}
\rotate
\tablecolumns{11}
\tablecaption{Upper limits of H$_2$ masses in the beam size \label{tbl3}}
\tablewidth{0pt}
\tablehead{
\colhead{Name} &\colhead{SFR$_{H\alpha}$} &\colhead{log(SFE)$_{H\alpha}$} &\colhead{SFR$_{NUV}$}&\colhead{log(SFE)$_{NUV}$} &\colhead{logL$_{CO}$} & \colhead{Beam Filling Factor} & \colhead{logM$_{H2_{Total}}$} & \colhead{M$_{H2_{Total}}$/M$_{HI}$}  &\colhead{logL$_{3.4}$} &\colhead{logM$_*$} \\
  &\colhead{M$_\odot$yr$^{-1}$} &\colhead{yr$^{-1}$}&\colhead{M$_\odot$yr$^{-1}$} &\colhead{yr$^{-1}$} &\colhead{Kkm$s^{-1}$pc$^2$}& &\colhead{M$_\odot$}& & \colhead{L$_\odot$} &\colhead{M$_\odot$} \\
}

\startdata
AGCNr188845 & 0.097  &-10.012 &         &         & $<$6.71  &  1.35  & $<$7.635 & $<$0.044  & 7.316 &8.154\\
AGCNr4528   & 0.079  &-10.500 &0.232  &-10.037    & $<$6.50  &  1.65  & $<$7.643 & $<$0.017 & 7.437 &8.290\\
AGCNr12212*\tablenotemark{a} & 0.111  &-10.300    &          &        & $<$5.747 & 1.55 & $<$7.612 &$<$0.0201 &   &           \\
            & 0.056  &-10.599 &         &         &          &          &          &          &        &           \\
AGCNr110150 & 0.108  &-10.457 &0.146   &-10.327   &$<$6.52   &  2.07   &$<$7.827 &$<$0.022 &           &           \\
AGCNr102243 & 0.829  &-9.865  &         &         &$<$7.10   &  1.35   & $<$8.017 & $<$0.017 & 7.665 &8.545\\
AGCNr12289*\tablenotemark{b}  & 0.339  &-10.776 &         &         &$<$7.79   &  1.00   & $<$8.615 & $<$0.020 & 8.795 &9.810\\
AGCNr102981*\tablenotemark{c} & NAN &      &     &  &  $<$6.411 &  1.76   & $<$7.07 & $<$0.018 & 7.760 &8.651\\
AGCNr102635 &0.367   &-10.085 &0.539   &-9.930    &$<$7.39   &  1.76   &$<$8.336  & $<$0.048 & 8.029 &8.952\\
NGC7589*\tablenotemark{d} & 0.173 &-10.782 &1.003 &-10.018   & $<$7.66 & 1.90   & $<$8.960 & $<$0.0806  & 8.894 &9.922\\
\enddata
\tablenotetext{a}{For AGCNr12212, the two different SFR$_{H\alpha}$
results are from the works\citep{2000AJ....119.2757V,2008MNRAS.390..466E}. We did not derive
its H$\alpha$ flux from our observation.}
\tablenotetext{b}{This just is the result with the observations of JCMT. The CO content of AGCNr12289 had been detected by \citet{2003ApJ...588..230O} with the IRAM and the ratio of M$_{H_2}$/M$_{HI}$ is 0.008.}
\tablenotetext{c}{For AGCNr102981, we did not derive its H$\alpha$ flux from our
  observation.}
\tablenotetext{d}{For NGC 7589, the SFR$_{H\alpha}$ is
calculated by SDSS spectra \citep{2016MNRAS.455.3148S}.}
\end{deluxetable}
\clearpage

\end{document}